\documentstyle[12pt,epsfig]{article}
\newcommand{\blankline}{\vskip .3cm}
\newcommand{\f}{\begin{equation}}
\newcommand{\ff}{\end{equation}}
\renewcommand{\thefootnote}{\fnsymbol{footnote}}
\def\be{\begin{equation}}
\def\ee{\end{equation}}

\def\ea{\end{eqnarray}}
\newcommand\pubnumber{SLAC-PUB-10158\\}
\newcommand\pubdate{\today}
\newcommand\hepnumber{hep-th/0309045}

\def\SLAC{Stanford Linear Accelerator Center and ITP\\
    Stanford University, Stanford, California 94309 USA}
\def\doeack{\footnote{Work supported by the Department of Energy,
                     contract DE--AC03--76SF00515.}}
\def\PI{Perimeter Institute for Theoretical Physics, Waterloo, Canada }
\def\UW{Department of Physics, University of Waterloo\\
Ontario, Canada }

\def\Title#1{\begin{center} {\Large #1 } \end{center}}
\def\Author#1{\begin{center}{ \sc #1} \end{center}}
\def\Address#1{\begin{center}{ \it #1} \end{center}}
\def\andauth{\begin{center}{and} \end{center}}

\newcommand\pubblock{\rightline{\begin{tabular}{l} \pubnumber\\
         \pubdate \\ \hepnumber \end{tabular}}}

\begin{document}
\begin{titlepage}
\pubblock

\vfill
\Title{Quantum Gravity and Inflation}
\vfill
\Author{Stephon Alexander\doeack}
\Address{\SLAC}
\medskip
\andauth
\medskip
\Author{Justin Malecki}
\Address{\UW}
\Address{\PI}
\medskip
\andauth
\medskip
\Author{Lee Smolin}
\Address{\PI}
\end{titlepage}
\def\thefootnote{\fnsymbol{footnote}}
\setcounter{footnote}{0}
\newpage

  \centerline{ABSTRACT}

Using the Ashtekar-Sen variables of loop quantum gravity, a new
class of exact solutions
to the equations of quantum cosmology is found for gravity
coupled to a scalar field that corresponds to inflating universes.
The scalar field, which has an arbitrary potential,  is treated as
a time variable, reducing the hamiltonian constraint to a
time-dependent Schroedinger equation. When reduced to the
homogeneous and isotropic case, this is solved exactly by a set
of solutions that extend the Kodama state, taking into account
the time dependence of the vacuum energy. Each quantum state corresponds 
to a classical solution of the Hamiltonian-Jacobi equation. The study of the
latter shows evidence for an attractor, suggesting a
universality in the phenomena of inflation.  Finally,
wavepackets can be constructed by superposing solutions with
different ratios of kinetic to potential scalar field energy,
resolving, at least in this case, the issue of normalizability
of the Kodama state.

 \blankline\blankline\blankline\blankline\blankline
 \blankline\blankline\blankline\blankline\blankline\blankline
 \blankline
emails:  salexander@itp.stanford.edu, jjmaleck@uwaterloo.ca,  lsmolin@perimeterinstitute.ca
 \eject
\tableofcontents
\newpage
\section{Introduction}
\bigskip

The inflationary scenario provides a framework for resolving the
problems of the standard big bang (SBB) and, most importantly,
provides a causal mechanism for generating structure in the
universe. Currently, however, a completely satisfactory realization of
inflation is still lacking. A hint for finding a concrete
realization of inflation comes from the transplankian problem.
Despite their successes in solving the formation of structure
problem, most scalar field driven inflation models generically predict
that the near scale invariant spectrum of quantum fluctuations
which seeded structure were generated in the transplanckian epoch.
This, however, is inconsistent with the assumptions of a weakly coupled
scalar field theory as well as the assumption that quantum gravity
effects can be ignored in the model~\cite{Martin}. One can then view the
transplankian problem as an indication, and hence an opportunity,
that the concrete derivation of inflation should be embedded in
quantum gravity.

There is also another interesting hint that suggests that quantum
gravity must play a role in our understanding of inflation.
Inflation addresses
the issue of initial conditions in the SBB, but solutions to scalar field
theory driven inflation suffer from geodesic
incompleteness. This is an indication that inflation itself
requires the specification of fine tuned initial conditions ~\cite{Mark}.
The issue of the robustness of the initial
conditions necessary to
start a phenomenologically acceptable period of inflation, and their
sensitivity to the
parameters of the underlying field theory, remain
open questions which should in principle be addressed by quantum
gravity.\footnote{There are other motivations for expecting a
quantum gravitational derivation of inflation but this is outside
the scope of this paper.  A good discussion of this issue is
nicely covered in a review by Robert Brandenberger
~\cite{BrandRev}.} Therefore, a major goal of quantum gravity and
cosmology is to find a quantum gravitational state which yields a
consistent description of inflation. If this is accomplished then
one may be in a better position to make observational predictions for
CMB experiments.

The problem of inflation in quantum gravity has been
much studied~\cite{HH,Hali,Andrei}. However in the past the
poor understanding of quantum gravity necessitated that the
study of inflation be restricted to
the semiclassical approximation. This restriction makes it difficult to
obtain reliable results about issues such as initial conditions and
transplankian effects that involve the regime in which quantum gravitational
effects will be significant.

In this light it is worth noting that in recent years a great deal of progress
has been made in a non-perturbative approach to
quantum gravity, called loop quantum gravity~\cite{lqg}.
It is then appropriate to investigate whether these advances
allow us to treat the problem of inflation within
cosmology more precisely. Recent results
of Bojowald and others~\cite{bojowald} indicate that in loop quantum gravity one can
find exact quantum states that allow us to investigate more precisely the
role of quantum gravitational effects on issues in cosmology, including inflation 
and the fate of the initial singularity.
Furthermore, for constant cosmological constant, there is an exact solution
to the quantum constraints that define the full quantum general relativity,
discovered by Kodama~\cite{Kodama}, which has both an exact Planck scale description and
a semiclassical interpretation in terms of deSitter spacetime.  While there are
open issues of interpretation concerning this 
state~\cite{witten-kodama,laurentlee, marugan}, it is also true that it can
be used as the basis of both non-perturbative and semiclassical 
caclulations~\cite{chopinlee,chopin,positive}.
Furthermore, exact results in the loop representation have made possible an understanding
of the temperature and entropy of deSitter spacetime~\cite{chopinlee,positive} in
terms of the kinematics of the quantum gravitational field.

Thus, there appears to be no longer any reason to restrict the study of quantum
cosmology to the semiclassical approximation. In this paper we provide more
evidence for this, by finding
 exact solutions to the equations of quantum
cosmology that provide exact quantum mechanical descriptions
of inflation.

In order to study the problem of inflation in quantum gravity we proceed
by several steps. First, we
couple general relativity to a scalar field, $\phi$, with an
arbitrarily chosen potential, $V(\phi )$.  We then choose a gauge
for the Hamiltonian constraints in which this scalar field is constant
on constant time hypersurfaces~\cite{gaugechoice}. 
This is appropriate for the study of inflation,
because it has been shown that, in terms of the standard cosmological time coordinates,
inflation cannot occur unless the fluctuations of the scalar field on constant time
surfaces are small~\cite{MWU}. There then always will exist a small, local rescaling
of time that makes the scalar field constant\footnote{Technical subtleties
regarding this choice of gauge are discussed below.}.

In this gauge the infinite number of Hamiltonian constraints are reduced to
a single, time dependent Shroedinger equation~\cite{gaugechoice}.  This is then solved, for
homogeneous, isotropic fields,  as follows. The corresponding classical
Hamilton evolution equations are solved exactly by a class of Hamilton-Jacobi functions.
Each solution involves the numerical integration of an ordinary
first order differential equation.  These reduce, in the limit
of vanishing slow role parameter, $\dot{V}/V$, to the Chern-Simons
invariant of the Ashtekar connection. This is good, as the latter is
known to be the Hamilton-Jacobi function for deSitter 
spacetime~\cite{Kodama,chopin,chopinlee,positive}.
By exponentiating the actions of these solutions, one obtains a semiclassical state
that reduces in the same limit to the Kodama state.  These new solutions
are only good in the semiclassical approximation. However, in this case it is
possible to find the corrections which make the wavefunctionals into exact
solutions of the time-dependent Schrodinger equation.

The connection to the Kodama state allows us also to investigate issues regarding
the physical interpretation of that state such as the normalizability of the
wavefuction~\cite{witten-kodama,laurentlee}. 
In the case studied here, each exact quantum state we find  is
delta-function normalizable in the physical Hilbert space,
corresponding to the reduced, homogeneous, isotropic degrees of freedom.
It is then interesting to ask whether fully normalizable states can be
constructed by superposing the different solutions. In fact,
at a given time, defined by the value of $\phi$,
the different solutions correspond to
different ratios of $\pi^2 / V(\phi )$, where $\pi $ is the
canonical momenta of the scalar field. It is then
reasonable to superpose such solutions as there is no reason
to believe that quantum state of the universe should at
early times be an eigenstate of the ratio of kinetic to
potential energy.
When we do this we find wavepackets which are exact normalizable solutions.
This suggests that the problem of normalizability of the Kodama state
in the exact theory may be resolved
similarly by adding matter to the theory and
then superposing extensions of the state corresponding to different eigenvalues
of the matter energy momentum tensors.

In the next section we describe the scalar field general relativity system in the
formalism of Ashtekar~\cite{lqg} together with the details of the procedure whereby the time
gauge is fixed.  Section 3 explains the reduction to homogeneous, isotropic fields
in these variables, while section 4 describes the solutions to the resulting classical
equations by means of a set of solutions to the Hamilton-Jacobi theory. The
solutions are studied numerically and evidence for an attractor is found.
In section 5 we quantize the homogeneous, isotropic system, discovering both
semiclassical and exact solutions to the Schroedinger equation.  Our conclusions
and some directions for further research are described in the final section.

\section{The Theory}

We consider general relativity coupled to a scalar field, $\phi$,  and
additional fields
$\Psi$, in the Ashtekar formulation of loop quantum 
gravity\footnote{For an introduction to the Ashtekar formalism in the 
context of cosmology,  see 
~\cite{positive}. Other good, more general and complete reviews are in 
~\cite{lqg}.}.  Working in the canonical formalism, the Hamiltonian constraint is of the form,
\f
{\cal H}= {\cal H}^{grav} + {1 \over 2} \pi^2 + {1 \over 2}E^{ai}E^b_i
\partial_a
\phi \partial_b \phi + {\cal H}^\Psi
\label{hamconstraint}
\ff
where $\pi$ is the conjugate momentum to $\phi$ and $E^{ai}$ is the conjugate momentum
to the complex $SO(3)$ connection $A_{ai}$.  The latter couple satisfy the
Poisson bracket relation
\f
\{A_{ai}( {\mathbf x} ), E^{bj}( {\mathbf y})\} =
i G \delta_b^a \delta_i^j \delta^3( {\mathbf x}, {\mathbf y})
\label{bracket}
\ff
where $G$ is Newton's constant.

In~(\ref{hamconstraint}) we use ${\cal H}^\Psi$ to denote the
Hamiltonian constraint for all other matter fields.  Unless otherwise noted,
we adopt the convention that lowercase latin indices $a, b, c, \ldots$ are
spatial indices while lowercase latin indices $i, j, k, \ldots$ are
internal $SO(3)$ indices.

We include the scalar field potential $V(\phi)$ in the gravitational term so that
\f
{\cal H}^{grav} = {1 \over l_p^2} \epsilon_{ijk} E^{ai} E^{bj} (F_{ab}^k +
{ G V(\phi ) \over 3}\epsilon_{abc}E^{ck} ).
\ff
where $F_{ab}^k$ is the curvature of the connection $A_{ai}$.  Note that any bare
cosmological constant, $\Lambda$, is included in $V(\phi)$ and $V(\phi) = \Lambda$ gives the Hamiltonian
constraint for general relativity sourced only by $\Lambda$ and no scalar field.

We also will impose the Gauss's law constraint, which enforces $SO(3)$
gauge invariance
\f
{\cal G}^i= D_a E^{ai}
\ff
and the diffeomorphism constraint, that imposes spatial
diffeomorphism invariance
\f
{\cal D}_a = E^{bi}F_{ab i} + \pi \partial_a \phi + {\cal D}^\Psi_a
\ff
where ${\cal D}^\Psi_a$ contains the matter fields.

We assume that spacetime has topology ${\cal M} = {\cal S} \times R$,
where ${\cal S}$ is the
spatial manifold.  As we are interested in cosmology we assume ${\cal S}$
has no boundary,
so that the Hamiltonian is given by
\f
H(N)= \int_{\cal S} N {\cal H}
\ff
which is defined for any lapse $N$. We note that $N$ has density weight
minus one.

\subsection{Fixing the Time Gauge}
\label{gaugefix}
In the Hamiltonian approach to general relativity, one is free to choose
any slicing of
spacetime into a one parameter family of spacelike surfaces, where that
parameter can be
considered to be a time coordinate. All such slicings are physically equivalent
and the Hamiltonian constraint generates
gauge transformations that take
us from any one spatial slice to any other.  The choice of a slicing is then a gauge choice.

In the usual treatments of inflation, the scalar field, $\phi$, is
required to be homogeneous
to a good approximation. As the deviations from homogeneity must be
small for inflation to occur
at all, in solutions to Einstein's equations in which inflation
takes place we can assume that
the surfaces of constant $\phi$ are spacelike.  The scalar field also varies as the universe
expands, that is, it is ``rolling down the hill''.  It is then possible to measure time during
inflation by the value of the $\phi$ field, keeping in mind that the
\emph{forward} progression of time corresponds to $\phi$ changing in the \emph{negative} direction.

As a result, we will choose to gauge fix the action of the Hamiltonian
constraint so
that $\phi$ is constant on constant time surfaces. We do this by
imposing the gauge condition~\cite{gaugechoice}
\f
\partial_a \phi =0.
\label{choice}
\ff
We need to ensure that this condition is maintained by evolution generated by
the Hamiltonian $H(N)$.  That is, we demand
\f
0= {d\partial_a \phi \over dt } = \{ \partial_a \phi , H(N) \} =
\partial_a(N \pi )
\ff
which tells us that, to ensure the gauge condition is preserved, we must
use a lapse
\f
N= k / \pi
\label{helping}
\ff
where $k$ is a constant.

The gauge condition (\ref{choice}) is not good on the whole configuration
space as
there are solutions to Einstein's equations for which none of the constant $\phi$
surfaces are spacelike.
Thus, (\ref{choice}) is more than a gauge choice, it is also a restriction
on the
space of solutions. Nevertheless, it is a restriction which is appropriate
to the study of
inflation as there are results~\cite{} that indicate that, in models where the 
metric is approximately
spatially homogeneous, inflation only takes
place for solutions
in which $\phi$ is also, to a good approximation, spatially homogeneous.

However, for most initial data that satisfies
(\ref{choice}), it is known that the gauge condition will not be preserved forever. The
condition
cannot be preserved if $\pi$ becomes zero at any point on ${\cal S}$.

Of course,
$\pi$ is chosen on the initial data surface, and then evolves.
Equation~(\ref{helping}) tells us that an infinite lapse is required to preserve the
gauge condition at points where $\pi$ vanishes.
So by fixing the gauge to (\ref{choice}) we will generally be able
to study only a finite period in the evolution of the universe. The extent
of the period in which the gauge choice is good depends on the initial
values taken for $\pi$ and the other fields. As we are interested in modeling
inflation in which deviations from homogeneity must be assumed to be small, we
will assume that the gauge choice remains good for the entire period of
inflation.
However, after we have built the quantum theory, we will have to be concerned with the
extent to which these conditions are reliable.

As the hamiltonain constraint does not commute with the gauge condition,
we have to solve all but one of the infinite number of Hamiltonian
constraints for the conjugate variable $\pi$. The one that is not solved is
the constraint whose lapse is inversely proportional to $\pi$, as that
constraint commutes with the gauge condition.

We then find that
\f
\pi = \pm [-2{\cal H}^{grav} - 2{\cal H}^\Psi ]^{1/2}.
\ff
There is one remaining Hamiltonian constraint which
must be imposed which is
\f
0= {\cal H}(N= k/\pi)= {k \over 2} \int_{\cal S} \pi - {1 \over 2}H
\label{lastconstr}
\ff
To get the dimensions right, $N$ should be dimensionless so we pick
$k=1/l_p^2$. Then
$H$ is the Hamiltonian for evolution in the gauge we have picked. It is
given by
\f
H= \pm { \sqrt{2} \over l_p^2} \int_{\cal S}[-{\cal H}^{grav} - {\cal H}^\Psi ]^{1/2}.
\ff
Finally, we define
\f
P= {1 \over l_p^2} \int_{\cal S} \pi
\ff
which has dimensions of energy.

Thus, if we call the time $T= l_p^2 \phi$, where the factor of $l_p^2$ is
included so that $T$ has dimensions of time, we have the Poisson bracket
\f
\{ T , P \} =1.
\ff
We then have from~(\ref{lastconstr})
\f
-P +H =0,
\label{newHamConstr}
\ff
taking note of the fact that the Hamiltonian is time dependent because
the potential term in ${\cal H}^{grav}$ depends on $T$.  Herein, we will
use $V(T)$ to denote the value of the original potential $V(\phi)$ evaluated at $\phi = T / l_p^2$.
We thus have
reduced general relativity to an ordinary Hamiltonian system with a time
dependent Hamiltonian.

\section{The Homogeneous Case}

In this paper we will be concerned with the spatially homogeneous case, in order to be able to
compare our approach to the standard results in inflationary cosmology. Thus,we
now turn to the reduction of the Hamiltonian system just derived to the case of
spatially homogeneous and isotropic universes. We will also consider from now on only
the case in which the scalar field is the sole matter field.

The description of deSitter spacetime in Ashtekar-Sen variables is described
in ~\cite{positive}. In a spatially flat slicing of deSitter spacetime, the
$SO(3)$ gauge can be chosen so that the solution is given by diagonal and
homogeneous fields
\f
A_{ai} = i\delta_{ai} A, \ \ \  E^{ai} = \delta^{ai} E
\label{homovar}
\ff
where $A$ and $E$ are constant on each spatial slice ${\cal S}$.

DeSitter spacetime with cosmological constant, $\Lambda$, is given by
\f
A= h f(t), \\\\\  E = f^2 , \\\\\ f(t) = e^{ht}
\ff
where the Hubble parameter is $h^2 = G \Lambda /3$ and $t$ is the usual time
coordinate defined so that the spacetime metric is given by
\f
ds^2 = -dt^2 + f^2 \left(ds_3\right)^2
\ff
where $\left(ds_3\right)^2$ is the flat metric on ${\cal S}$.

We will consider the generalization of deSitter spacetime in which the
homogeneous scalar field is used as the time coordinate, so that
$A$ and $E$ are separately functions of $T$.  In these coordinates,
the spacetime metric is
\f
ds^2 = -N^2 dT^2 + E(T) \left ( ds_3 \right )^2.
\label{metric}
\ff
where $N$ is the lapse (\ref{helping}).

The gauge and diffeomorphism
constraints are solved automatically by the reduction to a homogeneous solution
and the curvature is given by
\f
F_{abi } = - A^2 \epsilon_{abi}.
\ff
The gravitational part of the Hamiltonian constraint in this reduced model is given by
\f
{\cal H}^{grav} ={6 \over l_p^2} E^2 \left( -A^2 + { l_p^2 V(T) \over 3} E \right).
\ff

Given that ${\cal S}$ is not compact and our fields homogeneous, we must give a well
defined meaning to the integral over ${\cal S}$.  As space is homogeneous,
we can integrate over a compact region $\Sigma \subset {\cal S}$ such that
\f
\int_\Sigma =L^3
\ff
where $L$ is a fixed, non-dynamical length scale. In this way, $\Sigma$ is a
finite representative of the entire homogeneous space. We will use the
dimensionless ratio
$R=L/l_p$. $R$ is a free parameter in the homogeneous cosmological model
that is not
part of the full field theory, but arises from the reduction from
a field
theory to a mechanical system.

If there are no matter fields, we then have the Hamiltonian
\f
H(A, E, T) = \pm R^3 \sqrt{12 E^2 \left( A^2 - { l_p^2 V(T) \over 3} E \right) }.
\label{homoham}
\ff
In this way we have a finite dimensional Hamiltonian theory of cosmology with a spatially
homogeneous scalar field.

\section{Solution of the Hamiltonian System}
In order to find a solution for $A(T)$ and $E(T)$ we must first determine
their symplectic relationship.  Integrating $E^{bj}(\mathbf{y})$ over $\Sigma$
in~(\ref{bracket})\footnote{$E^{bj}$ has spatial density weight one and so can be
integrated over the spatial manifold.}, substituting our
homogeneous variables~(\ref{homovar}), and taking the trace over the free indices gives the Poisson bracket
\f
\{A, R^3 l_p E\} = 3
\label{newbracket}
\ff
where we recognize $(R^3 l_p / 3) E$ as the conjugate momentum to A.

\subsection{Derivation of the Hamilton-Jacobi Function}

We now proceed to solve our model using Hamilton-Jacobi (HJ) theory. We search for a
Hamilton-Jacobi
function $S(A, T)$ such that\footnote{Other approaches to inflation 
which involve solutions to the Hamilton-Jacobi equations, in the ``old"
canonical variables, are described in ~\cite{otherh-j}.}
\f
E= { 3 \over R^3 l_p } { \partial S \over \partial A} , \qquad
P = {\partial S \over \partial T }
\ff
where the normalization of E is due to the relationship~(\ref{newbracket}).
Substituting these into equation~(\ref{newHamConstr}) using the
Hamiltonian~(\ref{homoham}) gives the HJ equation
\f
{\partial S \over \partial T } = {6 \over l_p}
\sqrt{3 \left ({ \partial S \over \partial A} \right )^2 \left(A^2 -
{ l_p V(T) \over R^3 } { \partial S \over \partial A} \right) }
\label{hj1}
\ff
where we have chosen the positive root of the Hamiltonian.

A function $S_{cs}(A, T)$ will  have zero energy if it satisfies
\f
{\partial S_{cs} \over \partial A} = {R^3 \over l_p V(T)} A^2
\ff
so that $H(A, {\partial S_{cs} \over \partial A}, T) = 0$.  This implies
\f
S_{cs}(A, T) = {R^3 \over 3 l_p V(T)} A^3.
\label{likek}
\ff
However, this does not solve the Einstein equations because
\f
P = {\partial S_{cs} \over \partial T} = - {\dot{V} \over V} S_{cs} \neq 0
\ff
in general.

We pause in our derivation to note that (\ref{likek}) is related to the Kodama solution of the
full quantum theory ~\cite{Kodama}, because $S_{cs}$ is proportional to the Chern-Simons invariant
\f
\int Y_{CS}(A) = \int \textrm{Tr} ( A\wedge dA + {2
\over 3} A \wedge A \wedge A) = i R^3 l_p^3 A^3
\ff
where the last equality comes from using the homogeneous variables~(\ref{homovar}).

We can understand why $S_{cs}$ is not a solution to our model.  Were $V(T)$ constant so
that $\Lambda = V$ were the cosmological constant, (\ref{likek})
would be the Hamilton-Jacobi function for deSitter spacetime
(see ~\cite{positive}). Were this the case, we would
have to have $\pi = 0$ so that the scalar field contributed no kinetic
energy, but only the constant potential energy. As discussed above,
this would violate our gauge condition~(\ref{choice}).

The deviation
from deSitter spacetime is then given by terms
proportional to the ratio $ r= \pi^2 / 2 V $. This is proportional
to the ``slow roll parameter''
\f
\eta = l_p { \dot{V} \over V}.
\ff
When $r$ and $\eta$ are small, the kinetic energy of the scalar field is
small compared to its potential energy.

To get a solution to our model, which requires $\pi \neq 0$, we need to
modify deSitter spacetime by terms proportional to $r$ and $\eta$. To
do this, we modify the HJ function (\ref{likek}), which gives deSitter
spacetime, by adding a new dimensionless function of time, $u(T)$, so that,
\f
S_u (A,T) = { R^3 A^3 \over 3 l_p V } (1 + u(T)).
\label{happy}
\ff
We expect to find that $u(T)$ scales with $r$ and $\eta$. We plug this into the
Hamilton-Jacobi equation, (\ref{hj1},) and we find an equation for $u(T)$
\f
\dot{u} = {\dot{V} \over V}(1+u)+ {18 i \over l_p} (1+ u)\sqrt{3u}.
\label{ueq}
\ff
We note that the $A$ decouples, as each term in (\ref{hj1}) is proportional
to
$A^3$.  The result is that for every solution
to (\ref{ueq}) we get a cosmological model.

Before looking at the consequences of our solution, we first show that, for $u\neq 0$,
the spacetime deviates from deSitter spacetime. Recall that deSitter spacetime is the
unique Lorentzian solution which satisfies
the self-dual condition
\f
J_{ab}^i = F_{ab}^i + {l_p^2 \Lambda \over 3} \epsilon_{abc}E^{ci} =0
\ff
where the proportionality can be taken to define $\Lambda$.
In the homogeneous case we define $J$ such that $J_{ab}^i = J \epsilon_{ab}^i$ so the
self-dual condition reads
\f
J= -A^2 + {l_p^2 \Lambda \over 3} E=0.
\ff
For solutions generated by (\ref{happy}), taking $V(T) \equiv \Lambda$, we have
\f
J= u A^2
\ff
which, for non-trivial spacetimes, only vanishes if $u(T) \equiv 0$.

\subsection{Analysis of the Solutions}

Given our HJ equation~(\ref{happy}) we can calculate the lapse function~(\ref{helping}) to be
\f
N = - i {V l_p \over 6 A^3 (1 + u) \sqrt{3 u}}.
\label{lapse}
\ff
Introducing the integration constant $\alpha$ that one obtains from integrating
equation~(\ref{ueq}) and a further integration constant $\beta$ (both dimensionless)
we can derive a relationship between $A$ and $T$ by
\f
{\partial S_u \over \partial \alpha} = \beta.
\ff
This leads to the equation of motion
\f
A(T) = \left[ {3 l_p V(T) \beta \over R^3}
\left( {\partial u \over \partial \alpha} \right)^{-1} \right]^{1 \over 3}.
\label{AofT}
\ff
In practice, the values of $\alpha$ and $\beta$ are determined by the initial $A(T_0)$
and $E(T_0)$.  Note that the partial derivative of $u$ will, in general,
add further $T$ dependence to $A(T)$.  We can then use equation~(\ref{AofT}) to derive
the $T$ dependence of the conjugate momentum to be
\f
E(T) = {3^{5/3} \beta^{2/3} (1+u) \over R^2 l_p^{4/3} V^{1/3} }
\left({\partial u \over \partial \alpha} \right)^{-{2 \over 3}}
\label{EofT}
\ff

Given the form of the metric in these coordinates~(\ref{metric})
we see that the $TT$ component of the metric is given by $-N^2$.
In order for the metric to remain real and Lorentzian, we must have that
$N^2$ be positive and real.  Looking at~(\ref{lapse}) we see that this requires
that $u(T)$ be negative and real.  Furthermore, for the metric to remain Lorentzian
we require that $E > 0$.  Equation~(\ref{EofT}) then requires that $u>-1$.  Hence,
our gauge condition will break down unless
\f
-1 < u(T) < 0.
\label{physcond}
\ff
This then places a strong restriction on the value of $u$.

In order to get a sense of the behaviour of $u(T)$, we have
solved~(\ref{ueq}) numerically for the quartic potential 
\f V(T) = \lambda \left(T^2 - m^2 \right)^2 +
V_{min} \label{pot} 
\ff 
for several initial conditions $u(T_0)$
consistent with the restrictions discussed above.  
Potentials of this sort are renormalizable and satisfies the slow roll condition.

\begin{figure}
\begin{center}
\begin{tabular}{cc}
\epsfig{file=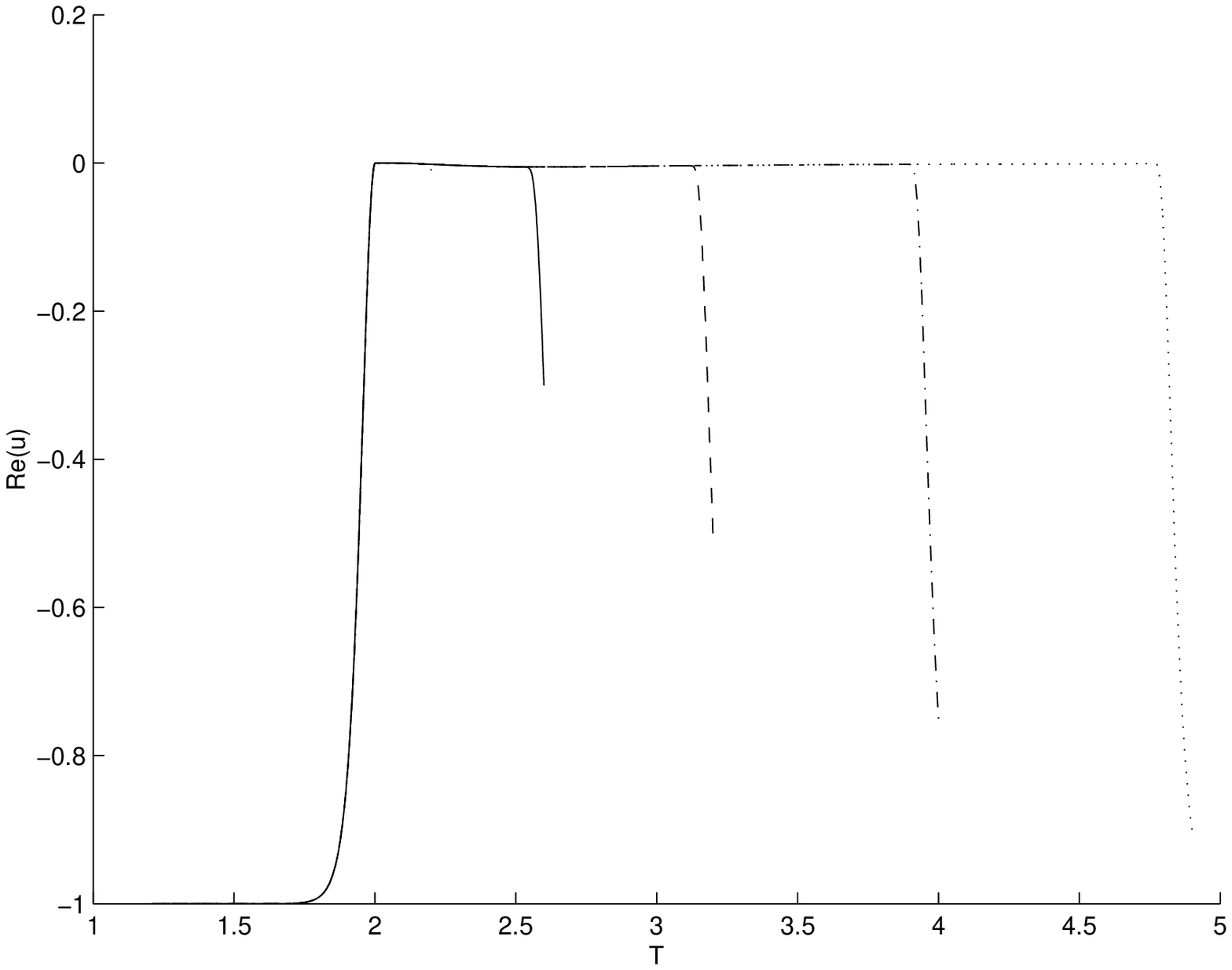, width=0.45\linewidth} &
\epsfig{file=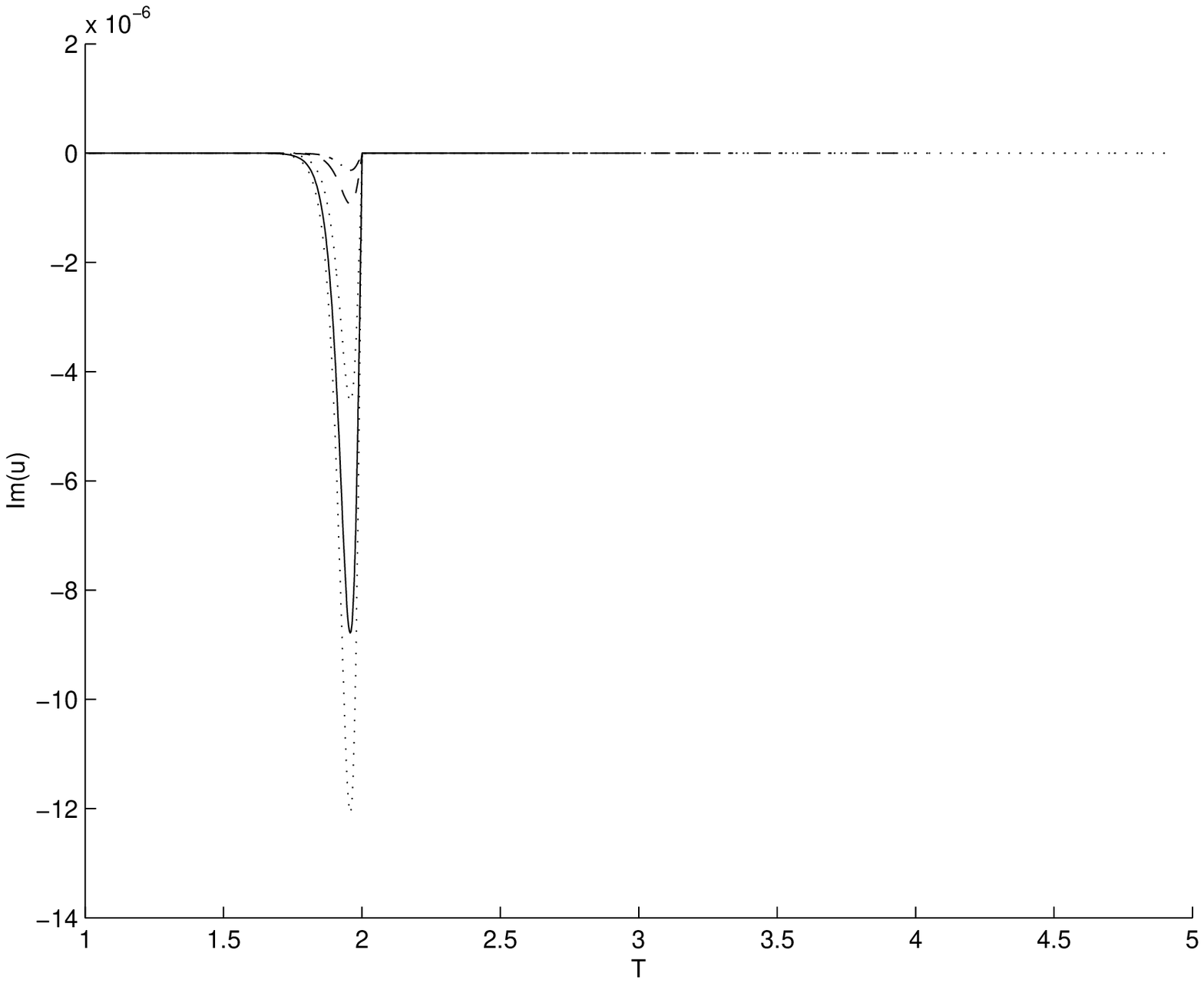, width=0.45\linewidth}
\end{tabular}
\end{center}
\caption{Numerical solutions of $u(T)$ using
the potential~(\protect\ref{pot}) for $\lambda = 1$, $m = 2$ and $V_{min} = 5$
and various initial conditions $T_0$, $u(T_0)$ satisfying $T_0 > 2$ and $-1 < u(T_0) < 0$.
The real part of $u$ is plotted on the left while the imaginary
part is on the right and the different line types on the left
and right correspond to the same solution.  The time variable
$T$ is in units of $l_p/c$.  Recall that \emph{forward} time
corresponds to $T$ evolving in the \emph{negative} direction.}
\label{fig:usoln}
\end{figure}

Unless otherwise noted, we always take $T_0 > m$ so that
$T = l_p^2 \phi$ proceeding in the negative direction corresponds to
$\phi$ ``rolling down the hill''.  These solutions can be seen in
Figure~\ref{fig:usoln} for $\lambda = 1$, $m = 2$ and $V_{min} = 5$. These plots
show the generic behaviour of $u$ for a wide range of parameters
in~(\ref{pot}).  

The most noticeable characteristic is the apparent attractor in
Figure~\ref{fig:usoln}.  That is, all initial conditions satisfying
our physical condition~(\ref{physcond}) merge to the same solution
for some $T < T_0$.  The significance of this apparent robustness to
initial conditions suggests a universality in the phenomena of inflation, which
should be further investigated.

Furthermore, in Figure~\ref{fig:usoln} we see that, for $T > m$, all of our
solutions remain within the bounds $-1 < u < 0$.  However, for $T < m$, $u$ quickly
becomes positive and takes on an imaginary component and the metric becomes unphysical.
$T = m$ corresponds to the scalar field having its minimum potential energy and is
traditionally the end of the inflationary period.  Hence, it is not surprising that
our gauge condition should break down at this point, as discussed in section~\ref{gaugefix}.
This break-down of the gauge condition at the minimum of the inflaton potential is a common
feature for all of the different parameters in~(\ref{pot}) that were attempted.  Of note
is the case where $V_{min} = 0$ in which the evolution equation~(\ref{ueq}) becomes singular for $T=m$.

While we defer a full numerical analysis of our model to a later time, we briefly discuss the
behavior of $u$ for ``unphysical'' initial conditions that do not satisfy~(\ref{physcond}).
It is clear from~(\ref{ueq}) that an initial condition $u(T_0) > 0$ forces $u$ to have an
imaginary component and hence give an unphysical metric.  For all of the initial conditions
$u(T_0) < -1$ that were attempted we found that $u$ rapidly diverged to $- \infty$.  Hence,
the physical conditions imposed on $u$ seem to be reflected in its functional behavior
governed by~(\ref{ueq}).

\section{Quantum Mini-Superspace}

We now proceed to build a quantum theory from our classical
Hamiltonian theory and find a full solution to the resulting Schr\"odinger equation.

First, we take quantum states to be functions $\Psi (A, T)$. We then
define $\hat{A}$ to be a multiplicative operator and define the operators
\f
\hat{P} \Psi = i \hbar {\partial \Psi \over \partial T} \ \ \ \
\hat{E} \Psi = - {i \hbar \over l_p R^3} {\partial \Psi \over \partial {A} }.
\label{ops}
\ff
Again, the normalization of $\hat{E}$ stems from the relation~(\ref{newbracket}).
Herein, we will work in units such that $\hbar = 1$.

The evolution equation becomes a time dependent Schr\"odinger equation,
\f
i {\partial \Psi \over \partial T} = \hat{H} \Psi
\label{schro}
\ff
where we choose the ordering of the quantum Hamiltonian to be
\f
\hat{H}= {\sqrt{12} i \over l_p} {\partial \over \partial A} \left(A \sqrt{\hat{J}} \right)
\ff
where we define $\hat{J}$ as
\f
\hat{J} := 1 - {i V l_p \over 3 R^3 A^2} {\partial \over \partial A}.
\ff
Note that we have taken the negative root of the classical Hamiltonian in constructing the Hamiltonian operator.  

We then take a semi-classical quantum state
\f
\Psi_u(A, T) = e^{i S_u(A, T)} =  e^{i {A^3 R^3 \over 3 V l_p} (1 + u)}
\label{psiu}
\ff
and note that it is an eigenfunction of $\hat{J}$ with a time dependent eigenvalue $-u(T)$
\f
\hat{J} \Psi_u = -u(T) \Psi_u
\ff
which implies
\f
\sqrt{\hat{J}} \Psi_u = i \sqrt{u(T)} \Psi_u.
\ff
Hence, the action of the Hamiltonian on our semi-classical state is
\f
\hat{H} \Psi_u =
\left[{6 i A^3 R^3 (1 + u) \sqrt{3 u} \over V l_p^2} + {6 \sqrt{3 u} \over l_p} \right] \Psi_u.
\label{hpsi}
\ff

We now show that $\Psi_u$ is, indeed, an approximate solution to the
Schr\"odinger equation~(\ref{schro}) by computing the time derivative
\f
{\partial \Psi_u \over \partial T} = - {6 A^3 R^3 (1 + u) \sqrt{3 u} \over V l_p^2} \Psi_u.
\label{dpsit}
\ff
where we have used the fact that $u(T)$ is a solution of~(\ref{ueq}).
Comparing~(\ref{hpsi}) and~(\ref{dpsit}) we see that, provided
\f
\left| {A^3 R^3 (1+u) \over V l_p} \right| \gg 1,
\ff
our semi-classical state~(\ref{psiu}) is indeed an approximate solution 
to the Schr\"odinger equation.

To find a full solution to~(\ref{schro}), we take the ansatz
\f
\Psi(A, T) = \Psi_u(A, T) \chi(T)
\label{ansatz}
\ff
where $\chi$ is an arbitrary function of $T$ alone.  Substitution of~(\ref{ansatz})
into the Schr\"odinger equation yields the equation
\f
{\partial \chi \over \partial T} = {6 i \sqrt{3u} \over l_p} \chi
\ff
which can be immediately integrated to give
\f
\chi(T) = e^{{6 \sqrt{3} i \over l_p} \int_{T_0}^T \sqrt{u(t)} dt}.
\ff
This yields
\f
\Psi(A, T) = e^{{6 \sqrt{3} \over l_p} \int_{T_0}^T \sqrt{-u(t)} dt + i{A^3 R^3 \over 3 V l_p} (1+u)}
\label{booyah}
\ff
which is a full solution to the Schr\"odinger equation~(\ref{schro}).
We have factored the $i$ into the square root term so that $\sqrt{-u}$ is real for solutions of $u$ that satisfy the physical condition~(\ref{physcond}).

To summarize, given a solution of~(\ref{ueq}) we have found a complete
solution~(\ref{booyah}) to the quantum theory of a homogeneous cosmology coupled
to a scalar field in a potential.

\subsection{Wavepackets and Normalizable States}

Finally, we describe the physical inner product and show how to construct
exact, normalizable, quantum states of the universe. In quantum gravity, the
physical inner product is
determined by the reality conditions for physical observables. In the
present case, in which all gauge degrees of freedom are fixed by either
gauge conditions or the reduction to homogeneous, isotropic solutions, $A(T)$ and $E(T)$ are
physical degrees of freedom, and they are indeed real.
In the representation we are using in which states are functionals of $A$
and $T$ and the latter is the time coordinate, the physical inner
product is hence,
\f
\left< \Psi (T) |\Psi (T) \right> = \int dA | \Psi (A,T) |^2.
\ff

Our exact solutions (\ref{booyah}) are phases, so long as $u(T)$ is real and negative,
corresponding to real solutions to the classical Einstein equations.
Hence, with this restriction the solutions are all delta function normalizable.

Following the usual procedure in quantum theory, we can construct normalizable
solutions by constructing wavepackets.  We may note that at a given $T$,
different classical solutions, with different values of $u(T)$ correspond to
cases in which the scalar field $\phi \approx T$ is moving at different rates.
Thus, while $V(T)$ is fixed at the same $T$, $\pi(T)$ can vary, leading
to different ratios of $\pi(T)^2/V(T)$.  However, quantum mechanically
we would expect that this ratio would not be a sharp observable. Thus the quantum state of
the expanding universe should not physically be built from a single solution
to the classical equations, $u(T)$. As there is no reason to expect that
the initial conditions for $u(T)$ are fixed classically in the very early
universe, we should expect the inflating universe to be described by a
a wavepacket corresponding to superposing over solutions that differ
in the value of $u(T)$ at fixed $T$.

To do this, let us fix $T=T_0$, and consider initial values for $u(T_0)$.
To construct a wavepacket we consider a central value $u_0$ and define
$v=u(T_0)-u_0$.  We then label
\f
\Psi(A, T_0)_v = e^{ i{A^3 R^3 \over 3 V(T_0) l_p} (1+u_0 + v)}
\label{booyah2}
\ff

Given a normalizable function $f(v)$ we may then define an exact solution
determined by the initial conditions
\f
\Psi(A, T_0)_f= \int dv f(v) \Psi(A, T_0)_v.
\ff
By a suitable choice of $f(v)$, with support in the physical interval,
$u=u_0+v \in (-1,0)$, the initial state is normalizable.

Let us then define ${\cal E}(T,v) =\int_{T_0}^T \sqrt{-u(t)} dt$
with initial condition $u(T_0)=u_0+v$ and $u_v(T)$ to be the value of
$u$ at time $T$ with the same initial condition.  Then the wavepacket at
later time is given by
\f
\Psi(A, T )_f = \int dv f(v)
e^{{6 \sqrt{3} \over l_p}{\cal E}(T,v) + i{A^3 R^3 \over 3 V l_p} (1+u_v(T))}.
\ff

\section{Conclusions and Discussion}

\label{discussion}

The results of this paper represent a step towards a detailed 
study of the very early universe beyond the semiclassical
approximation, in which quantum gravitational effects are treated in
a non-perturbative and background independent manner.
For each potential $V(\phi)$ and classical slow roll solution $u(T)$ consistent
with inflation, we have found a quantum state
given by (\ref{booyah}) which is an exact solution to the quantum
equations of motion, but has a classical limit given
by that classical solution. Furthermore, we can construct normalizable
states which are wavepackets around the initial conditions that generate
that classical solution. Thus, inflation is here described in terms
of exact quantum states.

As a byproduct, the simplicity of the Hamilton-Jacobi solutions to the
coupled Einstein-scalar field problem, using the Ashtekar formulation,
may provide a new, simplified approach to studying inflation classically.
In our first investigation of the problem we found an attractor, suggesting,
at least in this case, a possible
universality in the dynamics of the very early universe. This deserves more
investigation, as it may provide an understanding of the hypothesis of chaotic
inflation~\cite{Andrei}.

A number of very interesting questions remain, which these results suggest
can now be approached.

\begin{itemize}

\item{}It would be very interesting to understand the relationship of
these results to those of Bojowald and others, in the context of loop quantum
cosmology~\cite{bojowald}. In that case a similar reduction is used, and
solutions to quantum cosmology are found which are exact and non-perturbative.
However Bojowald's results obtained using a representation conjugate to that used here,
which is roughly a dimensionally
reduced spin network basis. A number of interesting results are obtained,
including indications that the initial singularity
is removed. It would then be very useful to establish a homogeneous version
of the loop transform, to express the states studied here in Bojowald's
representation.

\item{}The ordering of the Hamiltonian used here is not Hermitian.
It is easy in the dimensional reduction to find Hermitian orderings for
the Hamiltonian. The exact quantum states found here will solve the
Hermitian ordering of the Schrodeinger equation at the semiclassical
approximation.  It is challenging, but not impossible, to find exact
results for the case of Hermitian ordering.
If not, at least a semiclassical expansion could be constructed
that would be reliable above the Planck scale, whose leading order
terms would be given by the states found here.

\item{}It will be also interesting to incorporate the inhomogeneous modes
of the gravitational and matter fields by means of a
perturbative expansion around the states constructed here. The aim here
will be first principles predictions for transplankian effects in
the spectra of scalar and tensor
perturbations, as well as polarizations, detectible in cosmological
observations.

\item{}We note that the validity of our gauge condition ends
roughly when inflation ends, as surfaces of constant $\phi$ no longer
track surfaces of constant scale factor, once the universe enters the
stage where, in terms of the latter, the scalar field oscillates around
the miniumum of $V(\phi )$. We see that at this point in the classical
dynamics,
$u(T)$ becomes complex, leading to complex values of the
spacetime metric.
To study the problem of exiting from inflation it will then be
necessary to choose another time parameter, which is good throughout
the exit from inflation, and use the wavefunction generated here
as initial conditions for evolution in that parameter.

\item{}There exist extensions of the Kodama state to supergravity
with $N=1,2$~\cite{superkodama}.  It is then likely that the results of the present
paper can be extended to supersymmetric models of inflation.

\item{}It will be interesting to investigate whether the mechanism used
here to construct normalizable states will work in the full theory, perhaps
resolving the issue of the normalizability of the Kodama state.

\end{itemize}

\section*{ACKNOWLEDGEMENTS}

We are grateful to Robert Brandenberger, Laurent Freidel, John Moffat, 
Michael Peskin, and Hendryk Pfeiffer for conversations during the course of this work. LS would like to thank SLAC and
the Physics Department of Stanford University, and SA would like to thank
the Perimeter Institute for hospitality during the course of this work.

\end{document}